\documentclass[aps,pre,onecolumn,nofootinbib,11pt,tightenlines]{revtex4-1}
\usepackage{amsmath,amssymb,amstext,graphicx,color,xfrac,braket,accents,tabulary,booktabs,xcolor,subcaption}
\usepackage[unicode=true,bookmarks=true,
bookmarksnumbered=false,bookmarksopen=false,
breaklinks=false,
pdfborder={0 0 1},
backref=false,colorlinks=true]{hyperref}

\hypersetup{pdftitle={title}, pdfauthor={Ashmeet Singh, Sean M. Carroll}, citecolor=blue,linkcolor=blue,urlcolor=blue,citecolor=blue}

\usepackage{color}
\definecolor{purple}{rgb}{0.5,0,0.5}

\begin{document}

\newcommand{\be}{\begin{equation}}
\newcommand{\ee}{\end{equation}}
\newcommand{\bea}{\begin{eqnarray}}
\newcommand{\eea}{\end{eqnarray}}
\newcommand{\hst}{\widetilde{\mathcal{H}}} 
\newcommand{\iso}{\dot{=}}
\newcommand{\tdec}{{\{\theta\}}}
\newcommand{\pdec}{{\{\phi\}}}
\newcommand{\spsi}{\ket{\psi}}
\newcommand{\psimu}{\ket{\psi^{(\mu)}}}
\newcommand{\Cmu}{\left[ C^{(\mu)} \right]}
\newcommand{\Y}{\left[ C \right]}
\newcommand{\Ymean}{\left[ \bar{C} \right]}
\newcommand{\deltaY}{\left[ \Delta C \right]}
\newcommand{\OD}{\left[ O_{D} \right]}
\newcommand{\phiM}{\left[ \Phi \right]}
\newcommand{\Dim}{\textrm{dim\,}}
\newcommand{\Tr}{\textrm{Tr\,}}
\newcommand{\hs}{\mathcal{H}} 
\newcommand{\A}{\hat{A}}
\newcommand{\B}{\hat{B}}
\newcommand{\V}{\hat{V}}
\newcommand{\opphi}{\hat{\phi}}
\newcommand{\oppi}{\hat{\pi}}
\newcommand{\D}{\hat{D}}
\newcommand{\pos}{\hat{Q}}
\newcommand{\mom}{\hat{P}}
\newcommand{\ham}{\hat{H}}
\newcommand{\Tu}{\hat{T}_{u}}
\newcommand{\Tv}{\hat{T}_{v}}
\newcommand{\subA}{\mathcal{A}}
\newcommand{\subB}{\mathcal{B}}
\newcommand{\oprho}{\hat{\rho}}
\newcommand{\intham}{\hat{H}_{\rm{int}}}
\newcommand{\selfham}{\hat{H}_{\rm{self}}}
\newcommand{\eye}{\hat{\mathbb{I}}}
\newcommand{\M}{\hat{M}}

\newcommand{\sean}[1]{\textbf{\color{red}[#1]}}
\newcommand{\ashmeet}[1]{[\textbf{\color{blue}{#1}}]}

\baselineskip=14pt
\hfill CALT-TH-2018-29
\hfill

\vspace{0.5cm}
\thispagestyle{empty}

\title{Modeling Position and Momentum in Finite-Dimensional \\ Hilbert Spaces via Generalized Pauli Operators}
\author{Ashmeet Singh}
\email{ashmeet@caltech.edu}
\affiliation{Walter Burke Institute for Theoretical Physics, California Institute of Technology, Pasadena, CA 91125}
\author{Sean M. Carroll}
\email{seancarroll@gmail.com}
\affiliation{Walter Burke Institute for Theoretical Physics, California Institute of Technology, Pasadena, CA 91125}

\begin{abstract}
The finite entropy of black holes suggests that local regions of spacetime are described by finite-dimensional factors of Hilbert space, in contrast with the infinite-dimensional Hilbert spaces of quantum field theory.
With this in mind, we explore how to cast finite-dimensional quantum mechanics in a form that matches naturally onto the smooth case, especially the recovery of conjugate position/momentum variables, in the limit of large Hilbert-space dimension.
A natural tool for this task are the Generalized Pauli operators (GPO).
Based on an exponential form of Heisenberg's canonical commutation relation, the GPO offers a finite-dimensional generalization of conjugate variables without relying on any \textit{a priori} structure on Hilbert space. We highlight some features of the GPO, its importance in studying concepts such as spread induced by operators, and point out departures from infinite-dimensional results (possibly with a cutoff) that might play a crucial role in our understanding of quantum gravity. 
We introduce the concept of ``Operator Collimation," which characterizes how the action of an operator spreads a quantum state along conjugate directions.
We illustrate these concepts with a worked example of a finite-dimensional harmonic oscillator, demonstrating how the energy spectrum deviates from the familiar infinite-dimensional case.
\end{abstract}
\maketitle
\tableofcontents

\section{Introduction} 
\label{sec:GCA_intro}

The Hilbert space of a quantum field theory is infinite-dimensional, for three different reasons: wavelengths can be arbitrarily large, and they can be arbitrarily small, and at any one wavelength the occupation number of bosonic modes can be arbitrarily high.
Once we include gravity, however, all of these reasons come into question.
In the presence of a positive vacuum energy, the de~Sitter radius provides a natural infrared cutoff at long wavelengths; the Planck scale provides a natural ultraviolet cutoff at short wavelengths; and the Bekenstein bound \citep{bekenstein1973, bekenstein1981, bekenstein1994} (or more generally, black hole formation and its consequent finite entropy) provides an energy cutoff.
It therefore becomes natural to consider theories where Hilbert space, or at least the factor of Hilbert space describing our observable region of the cosmos, is finite-dimensional \citep{Banks:2000fe,Bao:2017rnv, Banks2000, Fischler2000, Witten:2001kn,Dyson2002,Parikh:2004wh,Carroll:2017kjo,Carroll:2018rhc}.

Our interest here is in how structures such as fields and spatial locality emerge in a locally finite-dimensional context.
Hilbert space is featureless: all Hilbert spaces of a specified finite dimension are isomorphic, and the algebra of observables is simply ``all Hermitian operators." 
Higher-level structures must therefore emerge from whatever additional data we are given, typically eigenstates and eigenvalues of the Hamiltonian and perhaps the amplitudes of a particular quantum state. (For work in this direction see \citep{giddings2015,Cotler:2017abq,Carroll:2018rhc,Giddings:2018koz}.)
One aspect of this emergence is the role of conjugate variables, generalizations of position and momentum.
To begin an exploration of how spacetime and locality can emerge from a Hamiltonian acting on states in a finite-dimensional Hilbert space, in this paper we consider the role of conjugate variables in a finite-dimensional context.

In the familiar (countably) infinite-dimensional case, such as non-relativistic quantum mechanics of a single particle, classical conjugate variables such as position ($q$) and momentum ($p$) are promoted to linear operators on Hilbert space obeying the Heisenberg canonical commutation relations (CCR),
\begin{equation}
  \label{GCA_CCR}
  \left[ \hat{q} , \hat{p} \right] = i \: ,
\end{equation}
where throughout this paper we take $\hbar = 1$.
In field theory, one takes the field and its conjugate momentum as operators labelled by spacetime points and generalizes the CCR to take a continuous form labelled by spacetime locations. \footnote{In making the transition to field theory, one transits from a finite to an infinite number of degrees of freedom and hence an uncountably infinite-dimensional Hilbert space (which is non-separable). In this case, there can be unitarily inequivalent representations of the CCRs, implying that the physical subspaces spanned by eigenstates
of operators in a particular representation will be different.  In Algebraic Quantum Field Theory
this is described by Haag's theorem\citep{haag55}.  Then different choices of states (a unit-normed, positive linear functional) on the algebra specify different inequivalent (cyclic) representations. }
\\

The Stone-von~Neumann theorem guarantees that there is a unique irreducible representation (up to unitary equivalence) of the CCR on infinite-dimensional Hilbert spaces that are \emph{separable} (possessing a countable dense subset), but also that the operators $\hat{q}, \hat{p}$ must be unbounded.
There are therefore no such representations on finite-dimensional spaces. 
\\

There is, however, a tool that works in finite-dimensional Hilbert spaces and maps onto conjugate variables in the infinite-dimensional limit: the Generalized Pauli operators (GPO).
As we shall see, the GPO is generated by a pair of normalized operators $\hat{A}$ and $\hat{B}$ -- sometimes written as ``clock'' and ``shift'' matrices -- that commute up to a dimension-dependent phase, 
\begin{equation}
\label{GCA_weylbraid}
\hat{A}\hat{B} = \omega^{-1}\hat{B}\hat{A} \: ,
\end{equation}
where $\omega = \exp\left(2 \pi i /N\right)$ is a primitive root of unity.
Any linear operator can be written as a sum of products of these generators.
Appropriate logarithms of these operators reduce, in the infinite-dimensional limit, to conjugate operators obeying the CCR.
The GPO therefore serves as a starting point for analyzing the quantum mechanics of finite-dimensional Hilbert spaces in a way that matches naturally onto the infinite-dimensional limit.

We will follow a series of papers by Jagannathan, Santhanam, Tekumalla and Vasudevan \citep{SanthanamTekumalla1976, Jagannathan1981, Jagannathan:1981ri} from the 1970-80's, which developed the subject of finite-dimensional quantum mechanics, motivated by the Weyl's exponential form of the CCR. These were introduced first by Sylvester \citep{465901} and then applied to quantum mechanics by von Neumann \citep{Neumann1931}, Weyl \citep{weyl1950theory}, and Schwinger \citep{Schwinger570,schwinger_book}, and since have been discussed by many others in various contexts, some representative papers include  \citep{doi:10.1119/1.1514208, 0034-4885-67-3-R03, Granik1996, Landi:1999ey, Jagannathan1983,1742-6596-538-1-012020,Varadarajan1995,Digernes_Husstad_Varadarajan_1999,doi:10.1142/S0129055X94000213,1751-8121-40-33-R01,STOVICEK1984157,Albeverio2000,MARCHIOLLI20121538,Marchiolli:2013fra,0295-5075-86-6-60005} (and references therein). This paradigm of the Generalized Pauli operators has also been referred in the vast literature on this subject as the discrete Heisenberg group or the finite Weyl-Heisenberg group and also has been called Schwinger bases, Weyl operators \citep{Bandyopadhyay2002} and generalized spin \citep{PITTENGER2004255}, among others.  From an algebraic point of view, this structure corresponds to a generalized Clifford algebra (GCA) \citep{Jagannathan:2010sb} with two generators that follow an ordered commutation relation.\footnote{In general, a GCA can be defined with more generators and their braiding relations. For instance, the Clifford algebra of the ``gamma" matrices used in spinor QFT and the Dirac equation is a particular GCA with 4 generators \citep{Jagannathan:2010sb}.} The basic mathematical constructions worked out in this paper are not new; our goal here is to distill the features of the GPO that are useful in the study of locally finite-dimensional Hilbert spaces in quantum gravity, especially the emergence of a classical limit.

The paper is organized as follows. 
In Section \ref{sec:GPO_conjugate}, we motivate the need for an intrinsic finite-dimensional construction by pointing out the incompatibility of conventional textbook quantum mechanics and QFT with a finite-dimensional Hilbert space; and follow it up by introducing and using the GPO to construct a finite-dimensional generalization of conjugate variables. 
In Section \ref{sec:schwingerlocality} we introduce the concept of Operator Collimation as a means to quantify and study the spread induced by operators along the conjugate variables. 
Section \ref{sec:finitedim} deals with understanding equations of motion for conjugate variables in a finite-dimensional context and how they map to Hamilton's equations in the large dimension limit, and we explore features of the finite-dimensional quantum mechanical oscillator. 

\section{Finite-Dimensional Conjugate Variables from the Generalized Pauli Operators}
\label{sec:GPO_conjugate}

\subsection{Prelude}
\label{sec:stonevonneumann}

Consider the problem of adapting the Heisenberg CCR (\ref{GCA_CCR}) to finite-dimensional Hilbert spaces.
One way of noticing an immediate obstacle is to take the trace of both sides; the left-hand side vanishes, while the right-hand side does not.
To remedy this, Weyl \citep{weyl1950theory} gave an equivalent version of Heisenberg's CCR in exponential form,
\begin{equation}
\label{GCA_weylCCR}
e^{i \eta \hat{p} } e^{i \zeta \hat{q} } = e^{i \eta \zeta } e^{i \zeta \hat{q} } e^{i \eta \hat{p} } \: ,
\end{equation}
for some real parameters $\eta$ and $\zeta$. 
This does indeed admit a finite-dimensional representation (which is unique, up to unitary equivalence, as guaranteed by Stone-von Neumann theorem which works for separable Hilbert spaces (hence, finite-dimensional and countable infinite-dimensional) \citep{Kronz2005}). 
One can interpret this to be a statement of how the conjugate operators $\hat{q}$ and $\hat{p}$ fail to commute, although the commutation relation $\left[ \hat{q} , \hat{p} \right]$ will now no longer have the simple form of a c-number (\ref{GCA_CCR}).

The GPO offers a natural implementation of Weyl's relation (\ref{GCA_weylCCR}) to define a set of intrinsically finite-dimensional conjugate operators.
In this section, we will follow the construction laid out in references \citep{SanthanamTekumalla1976, Jagannathan1981, Jagannathan:1981ri} in developing finite-dimensional quantum mechanics based on Weyl's exponential form of the CCR. 

\subsection{The Generalized Pauli Operators}

Consider a finite-dimensional Hilbert space $\hs$ of dimension 
\begin{equation}
\Dim \hs = N,
\end{equation} 
with $N < \infty$. Let us associate Generalized Pauli operators  (GPO) by equipping the space $\mathcal{L}(\hs)$ of linear operators acting on $\hs$ awith two unitary operators as generators of the group, call them $\hat{A}$ and $\hat{B}$, which satisfy the following commutation relation,
\begin{equation}
\hat{A}\hat{B} = \omega^{-1} \hat{B}\hat{A} \: ,
\end{equation}
where $\omega = \exp\left(2 \pi i /N\right)$ is a primitive root of unity. This is also known as the Weyl Braiding relation in the physics literature, and is the basic commutation relation obeyed by the generators. 
In addition to being unitary, the generators also satisfy the following toroidal property,
\begin{equation}
\label{GCA_AdBdI}
\hat{A}^{N} = \hat{B}^N = \eye \: ,
\end{equation} 
where $\eye$ is the identity operator on $\hs$. The spectrum of the operators is identical for both GPO generators $\hat{A}$ and $\hat{B}$,
\begin{equation}
\label{GCA_specAB}
\mathrm{spec}(\A) \: = \: \mathrm{spec}(\B) \: = \: \{1, \omega^{1},\cdots, \omega^{2}, \cdots, \omega^{N-1}  \} \: .
\end{equation}
Thus, in line with our Hilbert-space perspective, specifying just the dimension $N$ of Hilbert space is sufficient to construct the group, which determines the spectrum of the generators and the basic commutation relations. 

The GPO can be constructed for both even and odd values of  $N$ and both cases are important and useful in different contexts. In this section, let us specialize to the case of odd $N \equiv 2 l + 1$ for some $l \in \mathbb{Z}^{+}$, which will be useful in constructing conjugate variables whose eigenvalues can be thought of labelling lattice sites centered around 0. In the case of even dimensions $N = 2m$ for some $m \in \mathbb{Z}^{+}$, one will be able to define conjugate on a lattice labelled from $\{0,1,2,\cdots,N-1 \}$ and not on a lattice centered around $0$. For the case of $N = 2$, we recover the Pauli matrices, corresponding to $A = \sigma_{x}$ and $B = \sigma_{z}$.
Operators on {qubits} can be seen as a special $N=2$ case of the GPO. 
While the subsequent construction can be done in a basis-independent way, we choose a hybrid route, switching between an explicit representation of the GPO and abstract vector space relations, to explicitly point out the properties of the group. 

Let us follow the convention that all indices used in this section (for the case of odd $N = 2l+ 1$) for labelling states or matrix elements of an operator in some basis will run over 
\begin{equation}
i,j,k \ \in \ -l ,(-l + 1) ,\cdots,-1,0,1,\cdots, l-1, l\,. 
\end{equation}
The eigenspectrum of both GPO generators $\hat{A}$ and $\hat{B}$ can be relabelled as,
\begin{equation}
\mathrm{spec}(\A) \: = \: \mathrm{spec}(\B) \: = \: \{\omega^{-l}, \omega^{-l + 1},\cdots, \omega^{-1}, 1, \omega^{1}, \cdots, \omega^{l-1}, \omega^{l}  \} \: .
\end{equation}
There exists a unique irreducible representation (up to unitary equivalence) \citep{Jagannathan:2010sb} of the generators of the GPO defined via Eqs. (\ref{GCA_weylbraid}) and (\ref{GCA_AdBdI}) in terms of $N \times N$ matrices
 \begin{equation}
\label{GCA_Amatrix}
  A \: = \:      \begin{bmatrix}
       0  & 0  & 0 & \cdots  & 1          \\[0.3em]
        1  & 0  & 0 &  \cdots   & 0          \\[0.3em]
        0  & 1  & 0 &  \cdots   & 0          \\[0.3em]
       . & .  & \cdots   & .  & .        \\[0.3em]
              .  & .  & \cdots   & .    & .      \\[0.3em]
        0  & 0   & \cdots   & 1 & 0          \\[0.3em]
     \end{bmatrix}_{N \times N} \: ,
        \: \: \:
  B \: = \:      \begin{bmatrix}
       \omega^{-l}  & 0  & 0 & \cdots  & 0          \\[0.3em]
        0  & \omega^{-l+1}  & 0 &  \cdots   & 0          \\[0.3em]
       . & .  & \cdots   & .  & .        \\[0.3em]
              .  & .  & \cdots   & . & .          \\[0.3em]
        0  & 0  & 0 & \cdots   & \omega^{l}          \\[0.3em]
     \end{bmatrix}_{N \times N} \: .
\end{equation}
The $\hat{.}$ has been removed to stress that these matrices are representations of the operators $\A$ and $\B$ in a particular basis, in this case, the eigenbasis of $\B$ (so that $B$ is diagonal). More compactly, the matrix elements of operators $\A$ and $\B$ in this basis are,
\begin{equation}
\left[ A \right]_{jk} \equiv \braket{b_{j} | \A | b_{k}} =  \delta_{j,k+1} \: , \: \: \:
\left[ B \right]_{jk} \equiv \braket{b_{j} | \B | b_{k}}  = \omega^{j} \delta_{j,k} \: ,
\end{equation}
with the indices $j$ and $k$ running from $-l, \cdots , 0, \cdots, l$ and $\delta_{jk}$ is the Kronecker delta function. The generators obey the following trace condition,
\begin{equation}
\Tr \left(\A^{j}\right) = \Tr \left(\B^{j}\right) = N \delta_{j,0} \: .
\end{equation}

 Let us now further understand the properties of the eigenvectors of $\A$ and $\B$ and the action of the group elements on them.
 Consider the set $\{\ket{b_j}\}$ of eigenstates of $\hat{B}$, 
\begin{equation}
\label{GCA_Baction}
\hat{B} \ket{b_j} = \omega^j \ket{b_j} .
\end{equation}
As can be  seen in the matrix representation of $\A$ in Eq.~(\ref{GCA_Amatrix}), the operator $\hat{A}$ acts to generate cyclic shifts for the eigenstates of $\B$, mapping an eigenstate to the next,
\begin{equation}
\label{GCA_Aaction}
\hat{A}\ket{b_j} = \ket{b_{j+1}} \: .
\end{equation}
The unitary nature of these generators implies a cyclic structure in which one identifies $\ket{b_{l+1}} \equiv \ket{b_{-l}}$, so that $\hat{A}\ket{b_l} = \ket{b_{-l}}$. 

The operators $\A$ and $\B$ have the same relative action on the eigenstates of one another, as there is nothing in the group structure which distinguishes between the two. The operator $\B$ generates unit shifts in eigenstates of $\A$,
\begin{equation}
\label{GCA_BactionOnAstates}
\hat{B}\ket{a_k} = \ket{a_{k+1}} \: ,
\end{equation}
with cyclic identification $\ket{a_{l+1}} \equiv \ket{a_{-l}}$. Hence we have a set of operators that generate shifts in the eigenstates of the other, which is precisely the way in which conjugate variables act and which is why the GPO provides a natural structure to define conjugate variables on Hilbert space. While should think of these eigenstates of $\A$ and $\B$ to be marked by their eigenvalues on a lattice: there is no notion of a scale or physical distance at this point, just a lattice of states labelled by their eigenvalues in a finite-dimensional construction along with a pair of operators which translate each other's states by unit shifts, respectively. It should be mentioned at this stage that even though this construction lacks the notion of a physical scale, there still exists a \emph{symplectic structure} \citep{1751-8121-43-7-075303,Heydari:2015yqa} (and references therein). This is a rich topic with a lot of interesting details which we will not discuss here, and the interested reader is encouraged to look into the references mentioned above.

To further reinforce this conjugacy relation between $\A$ and $\B$, we see that they are connected to each under a {discrete Fourier transformation implemented by Sylvester's matrix $S$, which is an $N \times N$ unitary matrix connecting $A$ and $B$ via $S A S^{-1} = B$.
Sylvester's matrix in the $\{\ket{b_j}\}$ basis has the form $\left[ S \right] _{jk} = {\omega^{jk}}/{\sqrt{N}}$.
The GPO generators $\A$ and $\B$ have been studied in various contexts in quantum mechanics and are often referred to as ``clock and shift" matrices.
They offer a higher dimensional, non-hermitian generalization of the Pauli matrices.

The set of $N^2$ linearly independent unitary matrices $\{B^{b}A^{a} | b,a = -l , (-l+1),\cdots,0,\cdots,(l-1),l \}$, which includes the identity for $a = b = 0$, form a unitary basis for  $\mathcal{L}(\hs)$. Schwinger \citep{Schwinger570} studied the role of such unitary basis, hence this operator basis is often called {Schwinger unitary basis}. Any operator $\hat{M}  \in \mathcal{L}(\hs)$ can be expanded in this basis,
\begin{equation}
\label{Mexpansion}
\hat{M} = \sum_{b,a = -l}^{l} m_{ba} \B^b \A^a \: .
\end{equation}
Since from the structure of the GPO we have $\Tr  \left[ \left( \B^{b'} \A^{a'} \right)^{\dag}  \left( \B^{b} \A^{a}\right) \right]= N \: \delta_{b,b'} \delta_{a,a'}$, we can invert Eq.~(\ref{Mexpansion}) to get the coefficients $m_{ba}$ as,
\begin{equation}
\label{mba_coeff}
m_{ba} = \frac{1}{N} \Tr \left[ \A^{-a} \B^{-b} \hat{M} \right] \: .
\end{equation} 
Thus, in addition to playing the role of conjugate variables in a finite-dimensional construction, the GPO fits in naturally with the program of minimal quantum mechanics in Hilbert space \citep{Carroll:2018rhc, Giddings:2018koz} by being able to define a notion of conjugate variables, one is able to classify and use any other operator on this space, including the Hamiltonian that governs the dynamics. This notion will be important to us when we define the idea of conjugate spread of operators, the so-called ``Operator Collimation," in Section \ref{sec:schwingerlocality}.  

\subsection{Finite-Dimensional Conjugate Variables}

We are now prepared to define a notion of conjugate variables on a finite-dimensional Hilbert space. The defining notion for a pair of conjugate variables is identifying two self-adjoint operators that each {generate translations} in the eigenstates of the other. For instance, in textbook quantum mechanics, the momentum operator $\hat{p}$ generates translations in the eigenstates of its conjugate variable, the position operator $\hat{q}$, and vice-versa. Taking this as our defining criterion, we would like to define a pair of conjugate operators acting on a finite-dimensional Hilbert space, each of which is the generator of translations in the eigenstates of its conjugate. 

We define a pair $\opphi$ and $\oppi$ to be conjugate operators by making the following identification,
\begin{equation}
\label{GCA_phipidef}
\A \equiv \exp{(-i \alpha \oppi)} \: , \: \: \: \: \B \equiv \exp{(i \beta \opphi)} \: ,
\end{equation}
where $\alpha$ and $\beta$ are non-zero  real  parameters  which  set  the  scale  of  the  eigenspectrum  of  the
operators $\opphi$ and $\oppi$. These are bounded operators on $\mathcal{\hs}$, and due to the virtue of the GPO generators $\A$ and $\B$ being unitary, the conjugate operators $\opphi$ and $\oppi$ are self-adjoint, satisfying $\opphi^{\dag} = \opphi$ and $\oppi^{\dag} = \oppi$. The operator $\oppi$ is the generator of translations of $\opphi$ and vice-versa. 
The apparent asymmetry in the sign in the exponential in Eq. (\ref{GCA_phipidef}) when identifying $\opphi$ and $\oppi$ is to ensure that the $j$-th column (with $j = -l, -l+1, \cdots, 0, \cdots l-1, l$) of Sylvester's matrix $S$ that diagonalizes $A$ is an eigenstate of $\oppi$ with eigenvalue proportional to $j$, and hence on an ordered lattice.
Of course, $\opphi$ has common eigenstates with those of $\B$ and $\oppi$ shares eigenstates with $\A$. Let us label the eigenstates of $\opphi$ as $\ket{\phi_{j}}$ and those of $\oppi$ as $\ket{\pi_{j}}$ with the index $j$ running from $-l,\cdots,0,\cdots,l$. The corresponding eigenvalue equations for $\opphi$ and $\oppi$ can be easily deduced using Eqs. (\ref{GCA_phipidef}) and (\ref{GCA_specAB}),
\begin{equation}
\label{phi_pi_eig}
\opphi \ket{\phi_j} = j \left( \frac{ 2 \pi}{(2 l + 1) \beta} \right) \ket{\phi_j} \: ,
\: \: 
\oppi \ket{\pi_j} = j \left( \frac{ 2 \pi}{(2 l + 1) \alpha} \right) \ket{\pi_j}  \: .
\end{equation}

Let us now solve for the conjugate operators $\opphi$ and $\oppi$ explicitly by finding their matrix representations in the $\ket{\phi_{j}}$ basis. By virtue of being diagonal, the principle logarithm of $B$ is 
\begin{equation}
\label{GCA_lnB}
\log B = (\log \omega) \: \mathrm{diag} \left( -l, -l+1, \cdots, 0, \cdots, l-1, l \right) \: .
\end{equation} 
Hence we have the matrix representation of $\opphi$,
\begin{equation}
\label{GCA_phimatrix}
\braket{\phi_{j} | \opphi | \phi_{j'}} = j \left( \frac{ 2 \pi}{(2 l + 1) \beta} \right) \delta_{jj'} \: ,
\end{equation}
which is diagonal in the $\ket{\phi_{j}}$ basis as expected. To find a representation of $\oppi$ in this basis, we notice that $\A$ is diagonalized by Sylvester's matrix, hence we can get its principle logarithm as $\log A = S ^{-1} \left( \log B \right) S$.
In the case of odd dimension $N = 2 l + 1$, the principle logarithms of $A$ and $B$ are well-defined, and we are able to find explicit matrix representations for operators $\opphi$ and $\oppi$ as above. The conjugate operators $\opphi$ and $\oppi$ are connected through Sylvester's operator, 
\begin{equation}
\oppi = \left(\frac{- \beta}{\alpha}\right) \hat{S}^{-1} \opphi \hat{S} \: ,
\quad
\opphi = \left(\frac{- \alpha}{\beta} \right) \hat{S} \oppi \hat{S}^{-1} \: .
\end{equation}
The following parity relations are obeyed, since $S^{2}$ is the parity operator, $[S^{2}]_{jk} = \delta_{j,-k}$,
\begin{align}
\hat{S}^{4} &= \eye \: , \: \: \: \: \hat{S}^{2} \opphi \hat{S}^{-2} = - \opphi \: , \: \:  \: \: \hat{S}^{2} \oppi \hat{S}^{-2} = - \oppi \: .
\end{align}
These relations have the same form as in infinite-dimensional quantum mechanics. 

Using the expression $\log A  = S ^{-1} \left( \log B \right) S$, the matrix representation for $\oppi$ in the $\ket{\phi_{j}}$ basis is,
\begin{equation}
\label{GCA_pimatrix}
\braket{\phi_{j} | \oppi | \phi_{j'}} = \left( \frac{ 2 \pi}{(2 l + 1)^{2} \alpha} \right) \sum_{n = -l}^{l} n \exp{\left( \frac{2 \pi i  (j - j') n }{2l + 1} \right)} \: = \:
 \begin{cases}
        0 \: , &  \text{if } j = j' \\
        \\
       \left( \frac{ i \pi}{(2 l + 1) \alpha} \right) \text{cosec}{\left(\frac{2 \pi l (j - j')}{2l + 1}  \right)} \: , & \text{if } j \neq j'
        \end{cases}
\end{equation}
The eigenstates of both $\opphi$ and $\oppi$ each individually are orthonormal bases for the Hilbert space $\hs$,
\begin{align}
\braket{\phi_{j}|\phi_{j'}} = \delta_{j,j'} \: , \: \: \: 
\sum_{j = -l}^{l} \ket{\phi_{j}}\bra{\phi_{j }} = \eye \: , \: \: \: 
\braket{\pi_{j}|\pi_{j'}} = \delta_{j,j'} \: , \: \: \: 
\sum_{j = -l}^{l} \ket{\pi_{j}}\bra{\pi_{j }} = \eye \: .
\end{align}

Thus, using the generators of the GPO, we are able to naturally identify a notion of conjugate operators, each of which is the generator of translations for the eigenstates of the other as seen by Eqs. (\ref{GCA_phipidef}), (\ref{GCA_Aaction}) and (\ref{GCA_BactionOnAstates}).
While the GPO provides us with a notion of {dimensionless} conjugate variables that have familiar ``position/momentum" properties, there is no notion of a physical length scale as yet. 
The operators we will ultimately identify as classical position and momentum operators depend on a non-generic decomposition of Hilbert space into subsystems that makes emergent classicality manifest. This is the so called quantum factorization problem \citep{mereology,Tegmark:2014kka, Piazza:2005wm}, sometimes referred to as the set selection problem \citep{Dowker:1994dd}.

\subsection{The Commutator}

In this section, we will work out the commutation relation between conjugate operators $\opphi$ and $\oppi$ as defined from the GPO in a finite-dimensional Hilbert space and understand how they deviate from the usual Heisenberg CCR and converge to it in the large dimension limit. In the infinite limit, the conjugate operators $\opphi$ and $\oppi$ obey Heisenberg's form of the CCR $\left[ \opphi, \oppi \right] = i $,
while our conjugate variables based on Eq.~(\ref{GCA_phipidef}) satisfy the the GPO commutation relation,
\begin{equation}
\label{GPOcommutation2}
\exp{(-i \alpha \oppi)} \exp{(i \beta \opphi)} =  \exp{ \left(- \frac{2 \pi i}{2l + 1} \right) }  \exp{(i \beta \opphi)} \exp{(-i \alpha \oppi)}  \: .
\end{equation}
On expanding the left-hand side of the GPO braiding relation Eq.~(\ref{GPOcommutation2}) and using the Baker-Campbell-Hausdorff Lemma we obtain,
\begin{equation}
\label{weylexponential1}
\exp{\left(i \beta \opphi + [-i \alpha \oppi,i \beta \opphi] + \frac{1}{2!}[-i \alpha \oppi,[-i \alpha \oppi,i \beta \opphi]]  + \cdots \right)} \exp{(-i \alpha \oppi)}  = \exp{ \left( \frac{2 \pi i}{2l + 1} \right) }  \exp{(i \beta \opphi)} \exp{(-i \alpha \oppi)}  \: .
\end{equation}

While this holds for arbitrary real, non-zero $\alpha$ and $\beta$ for any dimension $N = 2l + 1$, let us focus on the infinite limit when $\opphi$ and $\oppi$ should satisfy Heisenberg's CCR of Eq.~(\ref{GCA_CCR}). Substituting this in Eq.~(\ref{weylexponential1}) we obtain,
\begin{equation}
\exp{\left(i \beta \opphi - i \alpha \beta \right)} \exp{(-i \alpha \oppi)}  = \exp{ \left(- \frac{2 \pi i}{2l + 1} \right) }  \exp{(i \beta \opphi)} \exp{(-i \alpha \oppi)} \: ,
\end{equation}
which immediately gives us a constraint on the parameters $\alpha$ and $\beta$,
\begin{equation}
\label{alphabeta_constraint}
  \alpha \beta = \frac{2 \pi}{2l+1} \: ,
\end{equation}
such that the commutation relation in the infinite-dimensional limit maps onto the Weyl form of the CCR, Eq.~(\ref{GPOcommutation2}). Thus, when Eq.~(\ref{alphabeta_constraint}) is satisfied, the commutator of $\opphi$ and $\oppi$ will obey Heisenberg's CCR in the infinite-dimensional limit. 

We will show this explicitly later in this section, but before that, let us first compute the commutator of $\opphi$ and $\oppi$ in finite dimensions. 
The matrix representation of $\left[ \opphi, \oppi \right]$ in the $\{\ket{\phi_{j}}\}$ basis is,
\begin{equation}
\label{commutator_constraint}
\braket{\phi_{j} | \left[ \opphi, \oppi \right] | \phi_{j'}} = \frac{4 \pi^2 (j - j')}{(2l + 1)^{3} \alpha \beta} \sum_{n = -l}^{l} n \exp{\left( \frac{2 \pi i (j - j') n }{2l + 1} \right)} =  \frac{2 \pi (j - j')}{(2l + 1)^{2} } \sum_{n = -l}^{l} n \exp{\left( \frac{2 \pi i (j - j') n }{2l + 1} \right)} \: .
\end{equation}
Imposing $\alpha \beta (2l + 1) = 2 \pi$ and performing the sum, the commutator becomes
\begin{equation}
\label{commutatorZ}
\braket{\phi_{j} | \left[ \opphi, \oppi \right] | \phi_{j'}}  = 
 \begin{cases}
        0 \: , &  \text{if } j = j' \\
        \\
       \frac{i \pi (j - j')}{(2l + 1) } \text{cosec}{\left(\frac{2 \pi l (j - j')}{2l + 1}  \right)}  \: , & \text{if } j \neq j'.
        \end{cases}
\end{equation}
Under the constraint of Eq. (\ref{alphabeta_constraint}), the matrix elements of $\opphi$ and $\oppi$ become,
\begin{equation}
\label{ph_pi_matrix_constraint}
\braket{\phi_{j} | \opphi | \phi_{j'}} = j \alpha \delta_{j,j'} \: ,
\: \: \: 
\braket{\phi_{j} | \oppi | \phi_{j'}} = \left( \frac{ \beta}{2 l + 1 } \right) \sum_{n = -l}^{l} n \exp{\left( \frac{2 \pi i  (j - j') n }{2l + 1} \right)} \: .
\end{equation}

While we need $\alpha$ and $\beta$ to satisfy Eq.~(\ref{alphabeta_constraint}) to obtain the correct limit of Heisenberg's CCR in infinite dimensions, there is still freedom to choose one of the two parameters independently. One possibility  is that their values are determined by the eigenvalues and functional dependence of the Hamiltonian on these conjugate operators. (Since powers of $\opphi$ and $\oppi$ generate the Schwinger unitary basis of Eq.~(\ref{Mexpansion}), any operator can be expressed as a function of these conjugate operators.) Alternatively, since there is no sense of scale at this level of construction and the conjugate operators are dimensionless and symmetric, one could by fiat impose $\alpha = \beta = \sqrt{2 \pi /(2 l + 1)}$ and accordingly change the explicit functional form of the Hamiltonian, which should have no bearing on the physics. 

The most important feature of the finite-dimensional commutator is its {non-centrality}, departing from being a commuting c-number (as it is in infinite dimensions). Many characteristic features of quantum mechanics and quantum field theory hinge on this property of a central commutator of conjugate operators. It is expected that the presence of a non-central commutator will induce characteristic changes in familiar results, such as computing the zero-point energy. Non-centrality allows for a richer structure in quantum mechanical models, as we will discuss in Section \ref{sec:finite_models}. 
Let us write down the finite-dimensional commutator as $\left[ \opphi,\oppi \right] = i \hat{Z}$, where $\hat{Z}$ is a hermitian operator whose matrix elements in the $\opphi$-basis can be read off from Eq. (\ref{commutatorZ}). The matrix $Z$ (i.e. the matrix elements of $\hat{Z}$ in the $\opphi$ basis) is a real, traceless (null entries on the diagonal), symmetric Toeplitz matrix. Such structure can be exploited to better understand deviations of the commutator in finite dimensions as compared to the usual infinite-dimensional results.

We now turn to recovering conventional notions associated with conjugate variables in quantum mechanics based on an infinite-dimensional Hilbert space. In the infinite-dimensional case of continuum quantum mechanics, we take $l \to \infty$ and at the same time make the spectral differences of $\opphi$ and $\oppi$ infinitesimally small so that they are now labelled by continuous indices on the real line $\mathbb{R}$, while at the same time respecting the constraint $\alpha \beta (2 l + 1) = 2 \pi$. While finite-dimensional Hilbert spaces in the $N\to\infty$ limit are not isomorphic to infinite-dimensional ones (even with countably finite dimensions), there is a way in which we can recover Heisenberg's CCR as $N \to \infty$. 

In the expression for the commutator in Eq.~(\ref{commutator_constraint}), replace $n/(2l+1)$ with a continuous variable $x \in \mathbb{R}$ and replace the sum with an integral with $dx \equiv 1/(2l+1)$ playing the role of the integration measure,
\begin{equation}
\braket{\phi_{j} | \left[ \opphi, \oppi \right] | \phi_{j'}} = 2 \pi (j - j') \int_{-\infty}^{\infty} dx\, x \exp{\left( 2 \pi i (j - j') x  \right)} \: .
\end{equation}
Since the labels $j$ and $j'$ are continuous, we can re-write the integral above as,
\begin{align}
\braket{\phi_{j} | \left[ \opphi, \oppi \right] | \phi_{j'}} &= 2 \pi (j - j') \frac{1}{2 \pi i} \frac{d}{d (j-j')} \int_{-\infty}^{\infty} dx  \exp{\left( 2 \pi i (j - j') x  \right)} \:  , \\
&= - i (j - j') \frac{d}{d (j-j')} \delta (j - j') \: , \\
&=  i  \delta (j - j') \: ,
\end{align}
where we have used $y \delta^{'}(y) = -\delta(y)$. Thus, we are able to recover Heisenberg's CCR as the infinite-dimensional limit of the Weyl braiding relation. It can be shown on similar lines that in the infinite-dimensional limit, $\oppi$ has the familiar representation of $-i d/d\phi$ in the $\opphi$ basis. Hence, finite-dimensional quantum mechanics based on the GPO reduces to known results in the infinite-dimensional limit, while at the same time offering more flexibility to tackle finite-dimensional problems, as might be the case for local spatial regions in quantum gravity. As we will discuss in Sections \ref{sec:finite_models} and \ref{sec:GCA_discussion}, infinite-dimensional quantum mechanics with cutoffs is very different from an intrinsic finite-dimensional theory; these difference could affect our understanding of fine-tuning problems due to radiative corrections, such as the hierarchy and cosmological-constant problems. Also, finite-dimensional constructions can offer new features in the spectrum of possible Hamiltonians, as we discuss in Section (\ref{sec:finite_models}).

\section{Operator Collimation: The Conjugate Spread of Operators}
\label{sec:schwingerlocality} 

The concept of locality manifests itself in different ways in conventional physics. In field theory, commutators of spacelike-separated fields vanish, the Hamiltonian can be written as a spatial integral of a Hamiltonian density $\hat{H} = \int d^{3}x\, \mathbf{\hat{H}}(\vec{x})$, and Lagrangians typically contain local interaction terms and kinetic terms constructed from low powers of the conjugate momenta.  Higher powers of the conjugate momenta are interpreted as non-local effects and are expected to be suppressed. Haag's formulation of algebraic QFT \citep{haag55, haagAQFT} is also based on an understanding of locality.\footnote{Localized states and their properties form an interesting set of ideas in quantum field theory (for example, see \citep{Pestun:2016zxk}) and have been debated on for a long time now. They connect to various important constructions and theorems such as Newton-Wigner \citep{RevModPhys.21.400,pittphilsci5427} localization and the Reeh-Schlieder theorem \citep{pittphilsci649,Halvorson:2000fy} in field theory. Such ideas will not be discussed here since our motivation is trying to understand emergence of structures such as spacetime, causality, and classicality from basic quantum mechanics, without presupposing any such structure.}

From a quantum information perspective, when we think about sub-systems in quantum mechanics as a tensor product structure in Hilbert space $\hs = \bigotimes_{j} \hs_{j}$, the interaction Hamiltonian is taking to be $k$-local on the graph \citep{Cotler:2017abq}, thereby connecting only $k$-tensor factors for some small integer $k$, thus reinforcing the local character of physical interactions. Typically, given a pair of conjugate variables, a dynamics worthy of the label ``local'' should have the feature that a state localized around a given position should not instantly evolve into a delocalized state.
For example, the Hamiltonian for a single non-relativistic particle typically takes the special form $\hat{H} \sim \hat{\vec{p}}^{2} / 2 + \hat{V}(\hat{\vec{x}})$ for classical conjugate variables of position $\vec{x}$ and momentum $\vec{p}$. Both the quadratic nature of the kinetic term and fact that the Hamiltonian is additively separable in the conjugate variables serve to enforce this kind of locality by not allowing arbitrarily large spread of localized position states.
 
In a theory with gravity, the role of locality is more subtle. 
On general grounds, considering the metric as a quantum operator (or as a field to be summed over in a path integral) makes it impossible to define local observables, since there is no unique way to associate given coordinate values with ``the same'' points of spacetime.
In the context of the black-hole information puzzle, the principles of holography (the number of degrees of freedom within a black hole scales as the area of the horizon) and horizon complementarity (infalling observers see degrees of freedom spread out according to principles of local quantum field theory, while external observers see them as scrambled across the horizon) strongly suggest that the fundamental degrees of freedom in quantum gravity are not locally distributed in any simple way \citep{'tHooft:1993gx, Susskind:1994vu, Donnelly:2018qya, Banerjee:2016mhh}.

En route to understanding how spacetime emerges from quantum mechanics, we would like to understand these features better in a finite-dimensional construction without imposing additional structure or implicit assumptions of a preferred decomposition of Hilbert space, preferred observables, or conventional locality. With this motivation in mind, we can consider an even more primitive notion of ``locality". The following notion of ``locality", which we will call Operator Collimation, is a purely Hilbert-space construction and does not depend or refer to any underlying causal structure, relativistic or otherwise. 
Within our framework of conjugate variables, this primitive kind of locality can be understood by studying how operators in general (and the Hamiltonian in particular) act to spread eigenstates of conjugate variables in Hilbert space.

In this section, we develop a notion of the conjugate spread of an operator. This quantity helps characterize the support of an operator along the two conjugate directions. While this notion is not intrinsically tied to any time evolution generated by a Hamiltonian, and rather discusses the how different operators have support with respect to the two conjugate variables, it can be adapted to connect with more conventional notions of locality once relevant structures such as space, preferred observables, classicality etc. have been emerged under the right conditions.

As discussed in Section~\ref{sec:GPO_conjugate}, the Schwinger unitary basis $\{B^{b}A^{a} | b,a = -l , (-l+1),\cdots,0,\cdots,(l-1),l \}$ offers a complete basis for linear operators in $\mathcal{L}(\hs)$. The GPO generator $\A$ corresponds to a unit shift in the eigenstates of $\opphi$, and $\B$ generates unit shifts in the eigenstates of $\oppi$; hence, a basis element  $B^{b}A^{a}$ generates $a$ units of shift in eigenstates of $\opphi$ and $b$ units in eigenstates of $\oppi$, respectively (up to overall phase factors). 

For more general operators, the shifts implemented by the GPO generators turn into spreading of the state.
Consider a self-adjoint operator $\M \in \mathcal{L}(\hs)$ expanded in terms of GPO generators,
\begin{equation}
\M = \sum_{b,a = l}^{l} m_{b,a} \B^{b} \A^{a} \: .
\end{equation}
Since $\M$ is self-adjoint $\M^{\dag} = \M$, we get a constraint on the expansion coefficients, $\omega^{-ba} m^{*}_{-b,-a} = m_{b,a}$, which implies $|m_{b,a}| = |m_{-b,-a}|$ since $\omega = \exp{\left( 2 \pi i/(2l+1) \right)}$ is a primitive root of unity. 
The coefficients $m_{b,a}$ are a set of basis-independent numbers that quantify the spread induced by the operator $\hat{M}$ along each of the conjugate variables $\opphi$ and $\oppi$. To be precise, $|m_{b,a}|$ represents the amplitude of $b$ shifts along $\oppi$ for an eigenstate of $\oppi$ and $a$ shifts along $\opphi$ for an eigenstate of $\opphi$ . The indices of $m_{b,a}$ run from $-l, \cdots, 0, \cdots, l$ along both conjugate variables and thus, characterize shifts in both increasing $(a$ or  $b > 0)$ and decreasing $(a$ or  $b < 0)$ eigenvalues on the cyclic lattice. The action of $\hat{M}$ on a state depends on details of the state, and in general will lead to a superposition in the eigenstates of the chosen conjugate variable as our basis states, but the set of numbers $m_{b,a}$ quantify the spread along conjugate directions by the operator $\hat{M}$ independent of the choice of state. The coefficient $m_{00}$ accompanies the identity $\eye$, and hence corresponds to no shift in either of the conjugate variables. 

From $m_{b,a}$, which encodes amplitudes of shifts in both $\opphi$ and $\oppi$ eigenstates, we would like to extract profiles which illustrate the spreading features of $\M$ in each conjugate variable {separately}. Since the coefficients $m_{b,a}$ depend on details of $\M$, in particular its norm, we define normalized amplitudes $\tilde{m}_{b,a}$ for these shifts,
\begin{equation}
\label{mba_norm}
\tilde{m}_{b,a} = \frac{m_{b,a}}{\sum_{b',a' = -l}^{l} |m_{b',a'}|} \: .
\end{equation}
Then we define the {$\opphi$-shift profile} of $\M$ by marginalizing over all possible shifts in $\oppi$,
\begin{equation}
\label{ushiftham}
m^{(\phi)}_{a} \: = \: \sum_{b = -l}^{l} |\tilde{m}_{b,a}| \: = \: \frac{\sum_{b = -l}^{l} |m_{b,a}|}{\sum_{b',a' = -l}^{l} |m_{b',a'}|}  \: ,
\end{equation}
which is a set of $(2l+1)$ positive numbers characterizing the relative importance of $\M$ spreading the $\opphi$ variable by $a$ units, $a = -l,\cdots,0,\cdots,l$. Thus, $\M$ acting on an eigenstate of  $\opphi$, say $\ket{\phi = j}$, will in general, result in a superposition over the support of the basis of the $\opphi$ eigenstates $\{ \ket{\phi = j + a \: (\mathrm{mod} \: l)} \}  \: \: \forall \: a$, such that the relative importance (absolute value of the coefficients in the superposition) of each such term is upper bounded by $m^{(\phi)}_{a}$.
 
Let us now quantify this spread by defining \emph{operator collimations} for each conjugate variable. Consider the {$\phi$-shift profile} first. Operators with a large $m^{(\phi)}_a$ for small $|a|$ will have small spread in the $\opphi$-direction, while those with larger $m^{(\phi)}_{a}$ for larger $|a|$ can be thought of connecting states further out on the lattice for each eigenstate. Following this motivation, we define the {$\phi$-collimation} $C_{\phi}$ of the operator $\M$ as,
\begin{equation}
\label{Su}
C_{\phi}(\M) = \sum_{a =-l}^{l} m^{(\phi)}_{a} \exp{\left(- \frac{|a|}{2l + 1} \right)} \: .
\end{equation}
The exponential function suppresses the contribution of large shifts in our definition of collimation. There is some freedom in our choice of the decay function in our definition of operator collimation, and using an exponential function as in Eq. (\ref{Su}) is one such choice. Thus, an operator with a larger $C_{\phi}$ is highly collimated in the $\opphi$-direction and does not spread out eigenstates with support on a large number of basis states on the lattice. 

On similar lines, one can define the {$\pi$-shift profile} for $\M$ as,
\begin{equation}
\label{vshiftham}
m^{(\pi)}_{b} \: = \: \sum_{a = -l}^{l} |\tilde{m}_{b,a}| \: = \: \frac{\sum_{a = -l}^{l} |m_{b,a}|}{\sum_{b',a' = -l}^{l} |m_{b',a'}|}  \: ,
\end{equation}
and a corresponding {$\pi$-collimation} $C_{\pi}$ with a similar interpretation as the $\opphi$-case,
\begin{equation}
\label{Sv}
C_{\pi}(\M) = \sum_{b =-l}^{l} m^{(\pi)}_{b} \exp{\left(- \frac{|b|}{2l + 1} \right)} \: .
\end{equation}
Operators such as $\M(\oppi)$ that depend on only one of the conjugate variables will only induce spread in the $\opphi$ direction since they have $m_{b,a} = m_{0,a} \delta_{b,0}$, hence they possess maximum $\pi$-collimation, $C_{\pi}(\M) = 1$, as they do not spread eigenstates of $\oppi$ at all. Having a large contribution from terms such as $m_{0,0}, m_{b,0}, m_{0,a}$ will ensure larger operator collimation, since there are conjugate direction(s) where the operator has trivial action and does not spread the relevant eigenstates.
 
 \begin{figure}[t]
\includegraphics[width=\textwidth]{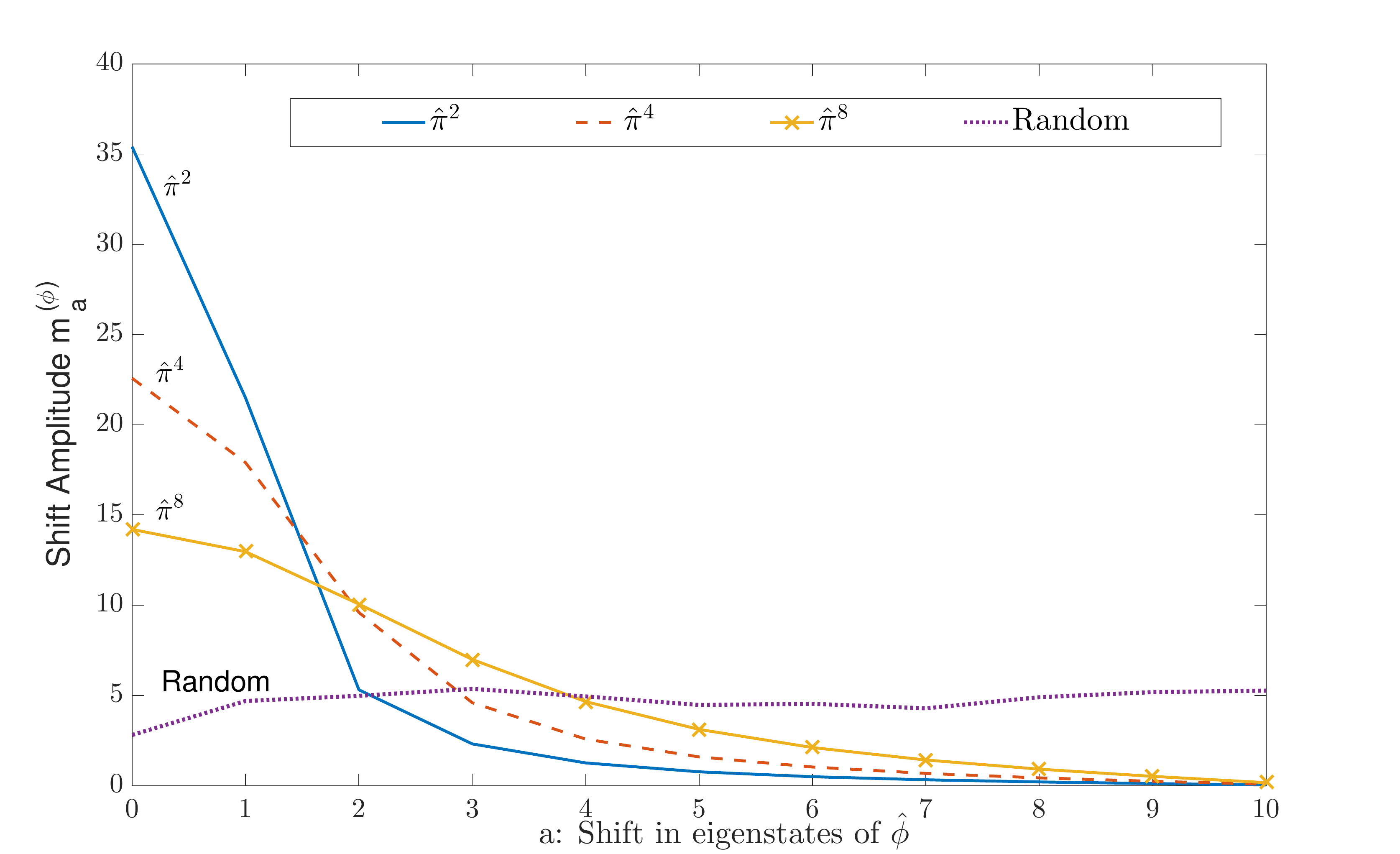}
\caption{Plot showing $\opphi$-shift profiles of various powers of $\oppi$. The quadratic operator $\oppi^{2}$ is seen to have the most collimated profile, implying that this operator does the least to spread the state in the conjugate direction. Also plotted is the profile for a random hermitian operator, for which the spread is approximately uniform.}
\label{schwinger_profiles}
\end{figure}

In general, we expect that operators which are additively separable in their arguments, $\M(\opphi,\oppi) = \M_{\phi}(\oppi) + \M_{\pi}(\pi)$, will have higher operator collimation as compared to a generic non-separable $\M$. Let us focus on operators depending only on one conjugate variable, say $\M \equiv \M(\oppi)$. While the maximum value of $C_{\pi}(\M(\oppi))$ can be at most unity, one can easily see that the hermitian operator,
\begin{equation}
\M(\oppi) = \frac{A + A^{\dag}}{2} = \frac{\exp{\left(- i \alpha \oppi\right)} + \exp{\left(i \alpha \oppi\right)}}{2} = \cos\left( \alpha \oppi \right) = \eye - \frac{\alpha^{2} \oppi^{2}}{2} + \frac{\alpha^{4} \oppi^{4}}{4} - \cdots \: ,\: ,
\end{equation}
has the least non-zero spread along the $\opphi$ direction: it connects only $\pm 1$ shifts along eigenstates of $\opphi$ and hence has highest (non-unity) $\phi$-collimation $C_{\phi}(\M)$. 
Thus, one can expect operators which are quadratic in conjugate variables are highly collimated. 
We see that the fact that real-world Hamiltonians include terms that are quadratic in the momentum variables (but typically not higher powers) helps explain the emergence of classicality: it is Hamiltonians of that form that have high operator collimation, and therefore induce minimal spread in the position variable. 

\begin{figure}[h]
\includegraphics[width=\textwidth]{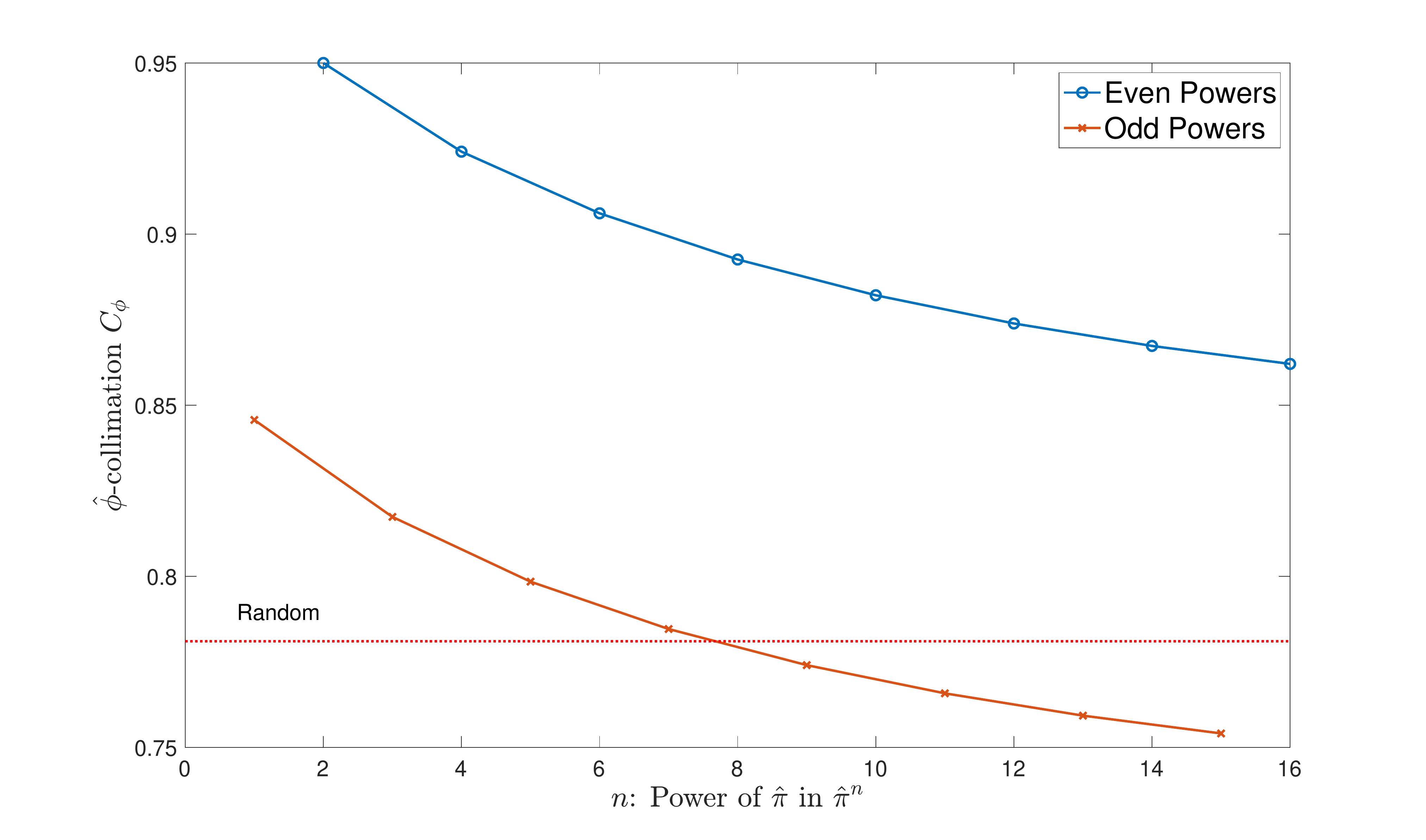}
\caption{$\phi$-collimation of various powers of $\oppi$. Even powers are seen to have systematically larger values of operator collimation. Also plotted for comparison is a line marking the $\phi$-collimation of a random hermitian operator.}
\label{schwinger_powers}
\end{figure}

Let us follow this idea further. The quadratic operator $\oppi^{2}$ has higher $\phi$-collimation than any other integer power $\oppi^{n}, n \geq 1 \: , \: n \neq 2$.  There is a difference between odd and even powers of $\oppi$, with even powers systematically having larger operator collimation than the odd powers. This is because odd powers of $\oppi$ no have support of the identity $\eye$ term in the Schwinger unitary basis expansion (and hence have $m_{00} = 0$), and having an identity contribution boosts collimation since it contributes to the highest weight in $C_{\phi}$ by virtue of causing no shifts. In Figure (\ref{schwinger_profiles}), we plot the $\phi$-shift profiles for a few powers of $\oppi$ and it is explicitly seen that quadratic $\oppi^{2}$ has the least spreading and hence is most $\opphi$-collimated, values for which are plotted in Figure (\ref{schwinger_powers}). Note that due to the symmetry $|m_{b,a}| = |m_{-b,-a}|$, we only needed to plot the positive half for $a > 0$, which captures all the information about the spread. Also, for comparison, we also plot the $\phi$-spread and its $\opphi$-collimation of a ``random" Hermitian operator (one whose matrix elements in the $\opphi$ basis are chosen randomly from a normal distribution); such operators spread states almost evenly and thus have low values of operator collimation. These constructions will be used to study locality properties of Hamiltonians and other observables in our upcoming work on studying emergent classical structure on Hilbert space \citep{mereology}.


\section{Finite-Dimensional Quantum Mechanics}
\label{sec:finitedim}

\subsection{Equations of Motion for Conjugate Variables}
\label{sec:finite_phasespace}

We would next like to understand equations of motion of conjugate variables defined by the GPO in a finite-dimensional Hilbert space evolving under a given Hamiltonian (and a continuous time parameter). In the large-dimension limit, and when appropriate classical structure has been identified on Hilbert space, conjugate variables $\opphi$ and $\oppi$ can be identified as position and momenta which satisfy Hamilton's equations of motion. As we will see, the structure of Hamilton's equations of motion is seen to emerge from basic algebraic constructions of the GPO when accompanied by an evolution by the Hamiltonian. Note that using the GPO, one can work with finite-dimensional phase space constructions, such as the discrete Wigner-Weyl construction, Gibbons-Hoffman-Wootters construction \citep{PhysRevA.70.062101,WOOTTERS19871}, and further discussed by various other authors \citep{PhysRevA.69.052112,1751-8121-50-50-504002,Werner2016,0305-4470-33-5-317,MARCHIOLLI20121538,0305-4470-33-36-307, 0305-4470-31-33-008,doi:10.1063/1.4961325} (and references therein). We will not discuss such finite-dimensional phase space ideas here but rather focus on understanding the equations of motion for conjugate variables and how they connect to Hamilton's equations.


Consider a Hamiltonian operator $\hat{H}^{\dag} = \hat{H}$ on $\hs$ which acts as the generator of time translations. We wish to construct operators corresponding to $``\partial H/\partial \phi"$ and $``\partial H/\partial \pi,"$ and be able to connect them with time derivatives of $\opphi$ and $\oppi$. 
 We saw that the operator $\A$ from the GPO generates translations in the eigenstates of $\opphi$, and $\B$ generates translations in eigenstates of $\oppi$. Notice that one can define a change in the $\phi$ variable as a finite central difference (we have used constraint $\alpha\beta = 2\pi (2l+1)$ from Eq. (\ref{alphabeta_constraint}) which gives the eigenvalues of $\opphi$ from Eq. (\ref{ph_pi_matrix_constraint})),
\begin{equation}
\label{delU}
\delta_{\phi}\opphi \equiv \left(\A^{\dag}\opphi\A - \A\opphi\A^{\dag} \right)  \: \implies \: \braket{\phi_{j} | \delta_{\phi}\opphi | \phi_{j'} } = 2 j  \alpha \delta_{j,j'} \: ,
\end{equation} 
up to ``edge" terms in the matrix where the finite-difference scheme will not act as in the usual way it does on a lattice due to the cyclic structure of the GPO eigenstates. Following this, we can write the change in $\ham$ due to a change in the $\phi$ variable (translation in $\opphi$) as a central difference given by,
\begin{equation}
\label{delH}
\delta_{\phi}\ham \: \equiv \: \left(\A^{\dag}\ham\A - \A\ham\A^{\dag}\right) \: .
\end{equation}
This allows us to define an operator corresponding to $\partial H/\partial \phi$ based on these finite central difference constructions,
\begin{equation} 
\label{delHdelUQM}
\hat{\left(\frac{\partial H}{\partial \phi}\right)}  = \frac{1}{2 \alpha} \: \left(\A^{\dag}\ham\A - \A\ham\A^{\dag} \right)\:  ,
\end{equation}
and similarly, for the change with respect to the other conjugate variable $\oppi$,
\begin{equation}
\label{delHdelVQM}
\hat{\left(\frac{\partial H}{\partial \pi}\right)} = \frac{1}{2 \beta} \left( \B^{\dag}\ham\B - \B \ham\B^{\dag} \right) \:  .
\end{equation}

The central difference is one possible construction of the finite derivative on a discrete lattice. One could use other finite difference schemes, but in the large-dimension limit, as we approach a continuous spectrum, any well-defined choice will converge to its continuum counterpart.

With this basic construction, let us now make contact with equations of motion (EOM) for the set of conjugate variables $\opphi$ and $\oppi$. We will work in the Heisenberg picture, where operators rather than states are time-dependent, even though we do not explicitly label our operators with a time argument. The Heisenberg equation of motion for an operator $\hat{\mathcal{O}}$ that is explicitly time-independent ($\partial_t\hat{\mathcal{O}}=0$) is,
\begin{equation}
\label{heisenbergEOM}
\frac{d}{dt}\hat{\mathcal{O}} = i \left[ \ham, \hat{\mathcal{O}} \right] \: .
\end{equation} 
In particular, for the time evolution of $\oppi$, we expand the right hand side of Eq.~(\ref{delHdelUQM}) using the Baker-Campbell-Hausdorff formula, and isolate the commutator $i \left[ \ham, \oppi \right]$ that will be the time rate of change of $\oppi$. One can easily show that, 
\begin{equation}
\label{Veom_finite}
\frac{d}{dt} \oppi = i \left[ \ham, \oppi \right]  = -  {\left(\frac{\partial H}{\partial \phi}\right)}_{op}  + \sum_{n = 3}^{\mathrm{odd}} \frac{i^n}{n!} \alpha^{n-1} \left[ \underline{\oppi} , \ham \right]_{n} \: ,
\end{equation}
where we have defined $\left[ \underline{\oppi} , \ham \right]_{n}$ as the $n$-point nested commutator in $\oppi$,
\begin{equation}
\label{npointcommutator}
\left[ \underline{\oppi} , \ham \right]_{n} = \left[\oppi,\left[\oppi,\left[\oppi \cdots \: \mathrm{(\textit{n}\ times)}, \ham \right]\cdots\right]\right] \: .
\end{equation}
The corresponding equation for $\opphi$ is likewise,
\begin{equation}
\label{Ueom_finite}
\frac{d}{dt} \opphi = i \left[ \ham, \opphi \right]  =  {\left(\frac{\partial H}{\partial \pi}\right)}_{op} + \sum_{n = 3}^{\mathrm{odd}} \frac{i^n}{n!} \beta^{n-1} \left[ \underline{\opphi} , \ham \right]_{n} \: .
\end{equation}
In the infinite-dimensional limit we take $l \to \infty$, and $\alpha$ and $\beta$ are taken to be infinitesimal but obeying $\alpha\beta (2l+1) = 2\pi$ to recover back the Heisenberg CCR. As expected, the equations of motion simplify to resemble Hamilton's equations of motion from classical mechanics,
\begin{equation}
\label{Veom_infinite}
\frac{d}{dt} \oppi = i \left[ \ham, \oppi \right]  = - {\left(\frac{\partial H}{\partial \phi}\right)}_{op} \: ,
\end{equation}
and
\begin{equation}
\label{Ueom_infinite}
\frac{d}{dt} \opphi = i \left[ \ham, \opphi \right]  =  {\left(\frac{\partial H}{\partial \pi}\right)}_{op} \: .
\end{equation}
These are intrinsically quantum equations for a set of conjugate variables from the GPO. They resemble the form of the classical equations of motion, but they do not necessarily describe quasiclassical dynamics. The emergence of quasiclassicality and identification of $\opphi$ and $\oppi$ with the classical conjugate variables of position and momentum is possible only in special cases when the substructure in Hilbert space allows for decoherence and robustness in the conjugate variables chosen. This is the concern of the quantum factorization problem of our upcoming work \citep{mereology}.

\subsection{The Finite-Dimensional Quantum Harmonic Oscillator}
\label{sec:finite_models}

With this technology of conjugate variables from the GPO, we can revisit some important models in quantum mechanics from a finite-dimensional perspective to compare the results with the usual infinite-dimensional results on $\mathbb{L}_{2}(\mathbb{R})$. All such results from finite-dimensional models will converge to the conventional infinite-dimensional ones when we take the limit $\Dim \hs \to \infty$. 

We will focus on a finite-dimensional version of the {harmonic oscillator}. Consider the following Hamiltonian $\ham$ operator for an oscillator  with ``frequency" $\Omega$ on a finite-dimensional Hilbert space $\hs$ with $\Dim \hs = 2l+ 1$, and let $\opphi$ and $\oppi$ be conjugate operators from the GPO,
\begin{equation}
\label{Hsho}
\ham = \frac{1}{2} \oppi^{2} + \frac{1}{2} \Omega^{2} \opphi^{2} \: = \Omega\left(\hat{a}^{\dag} \hat{a} + \frac{1}{2} \left[ \hat{a},\hat{a}^{\dag} \right] \right) \: .
\end{equation}
At this stage $\opphi$ and $\oppi$ are dimensionless operators, and $\Omega$ is a dimensionless parameter, so the Hamiltonian is also dimensionless. One can  define a change of variables,
\begin{align}
\hat{a} = \sqrt{\frac{\Omega}{2}} \opphi + \frac{i}{\sqrt{2 \Omega}} \oppi \: , \: \: \:
\hat{a}^{\dag} = \sqrt{\frac{\Omega}{2}} \opphi - \frac{i}{\sqrt{2 \Omega}} \oppi \: ,
\end{align}
but as we will see, these will \emph{not} serve as ladder or annihilation/creation operators in the finite-dimensional case,
since the non-central nature of the commutator carries through, $\left[ \opphi, \oppi \right] = i \left[ \hat{a},\hat{a}^{\dag} \right] = i \hat{Z}$.

 Due to finite-dimensionality of Hilbert space, and finite separation between eigenvalues of the conjugate variables, standard textbook results such as a uniformly spaced eigenspectrum will no longer hold. Depending on the interplay of eigenvalues of $\oppi$ and $\Omega \opphi$, there is an effective separation of scales, and correspondingly, the eigenvalue spectrum will have different features to reflect this. In the infinite-dimensional case, for any finite $\Omega$ the spectra of $\oppi$ and $\Omega \opphi$ match, since the conjugate operators have continuous, unbounded eigenvalues (the reals $\mathbb{R}$). In this sense, there is more room for non-trivial features in the finite-dimensional oscillator as compared to the infinite case. 
 
\begin{figure}[t]
\includegraphics[width=\textwidth]{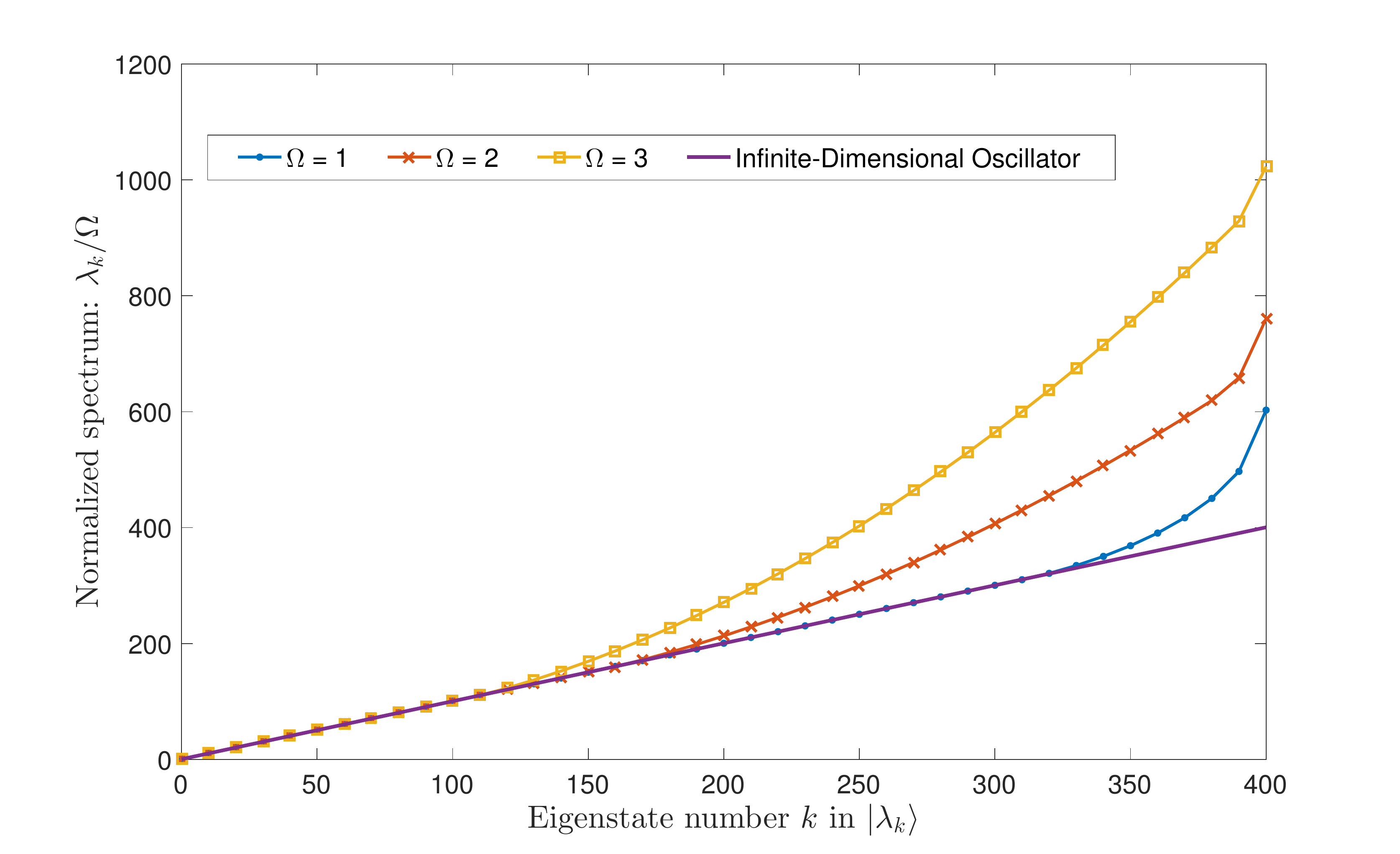}
\caption{Eigenspectrum (normalized by $\Omega$) for a $\Dim \hs = 401 (l  = 200)$ finite-dimensional oscillator for different values of $\Omega$. Depending on the value of $\Omega$, the spectrum deviates from the vanilla spectrum of the infinite-dimensional oscillator, which has also been plotted for comparison.}
\label{oscillator_spec}
\end{figure}

 In the eigenbasis of $\opphi$, the matrix elements of the Hamiltonian are,
\begin{equation}
\left[ \ham \right]_{jj'}  = 
 \begin{cases}
        \sum_{n \neq j} \frac{\pi}{4(2l+1)} \csc^{2}{\left( \frac{2 \pi l}{2l + 1} (j - n) \right)} + \frac{\Omega^{2} \pi}{2l + 1} j^{2}  \: , &  \text{if } j = j' \\
        \\
    \sum_{n \neq j, n \neq j'} \frac{\pi}{4(2l+1)} \csc{\left( \frac{2 \pi l}{2l + 1} (j - n) \right)} \csc{\left( \frac{2 \pi l}{2l + 1} (n - j') \right)}    \: , & \text{if } j \neq j'
        \end{cases}
\end{equation}
where we have used the constraint $\alpha = \beta = \sqrt{2 \pi / 2l+1)}$ as described in Section (\ref{sec:GPO_conjugate}), and all sums and indices run from $-l,\cdots,0,\cdots,l$. In the infinite-dimensional case, one can solve for the spectrum of the harmonic oscillator and obtain equispaced eigenvalues, which we refer to as the ``vanilla'' spectrum,
\begin{equation}
\lambda^{(\mathrm{vanilla})}_{n} = \left( n + \frac{1}{2} \right) \Omega \:  ,  n = 0,1,2,\cdots \: .
\end{equation}
The finite-dimensional case is more involved and we were unable to find an analytic, closed form for the spectrum $\{ \lambda_{k} \}$ in terms of $l$ and $\Omega$. We can solve for the spectrum numerically for different values of $l$ and $\Omega$, and here we point out a few important features.  
 
First consider the spectra of various oscillators with different $\Omega$ and how they compare with the vanilla, infinite-dimensional spectrum. In Figure (\ref{oscillator_spec}), we plot the spectrum for a $\Dim \hs = 401$ ($l  = 200$) finite-dimensional oscillator for different values of $\Omega$. Depending on how much $\Omega$ breaks the symmetry between eigenstates of $\oppi$ and $\Omega \opphi $ (corresponding to $\mathrm{max}(\Omega,1/ \Omega)$), the spectrum of the finite oscillator deviates from the vanilla, infinite-dimensional case and is no longer uniformly spaced. For the lower eigenvalues (what constitutes ``lower" depends on $\Omega$), both spectra match, and for larger eigenvalues, the finite-dimensional oscillator is seen to have larger values as compared to the vanilla case. On the same figure, we have also plotted part of the equispaced vanilla spectrum (which holds in infinite dimensions) for comparison.
\begin{figure}[t]
\includegraphics[width=\textwidth]{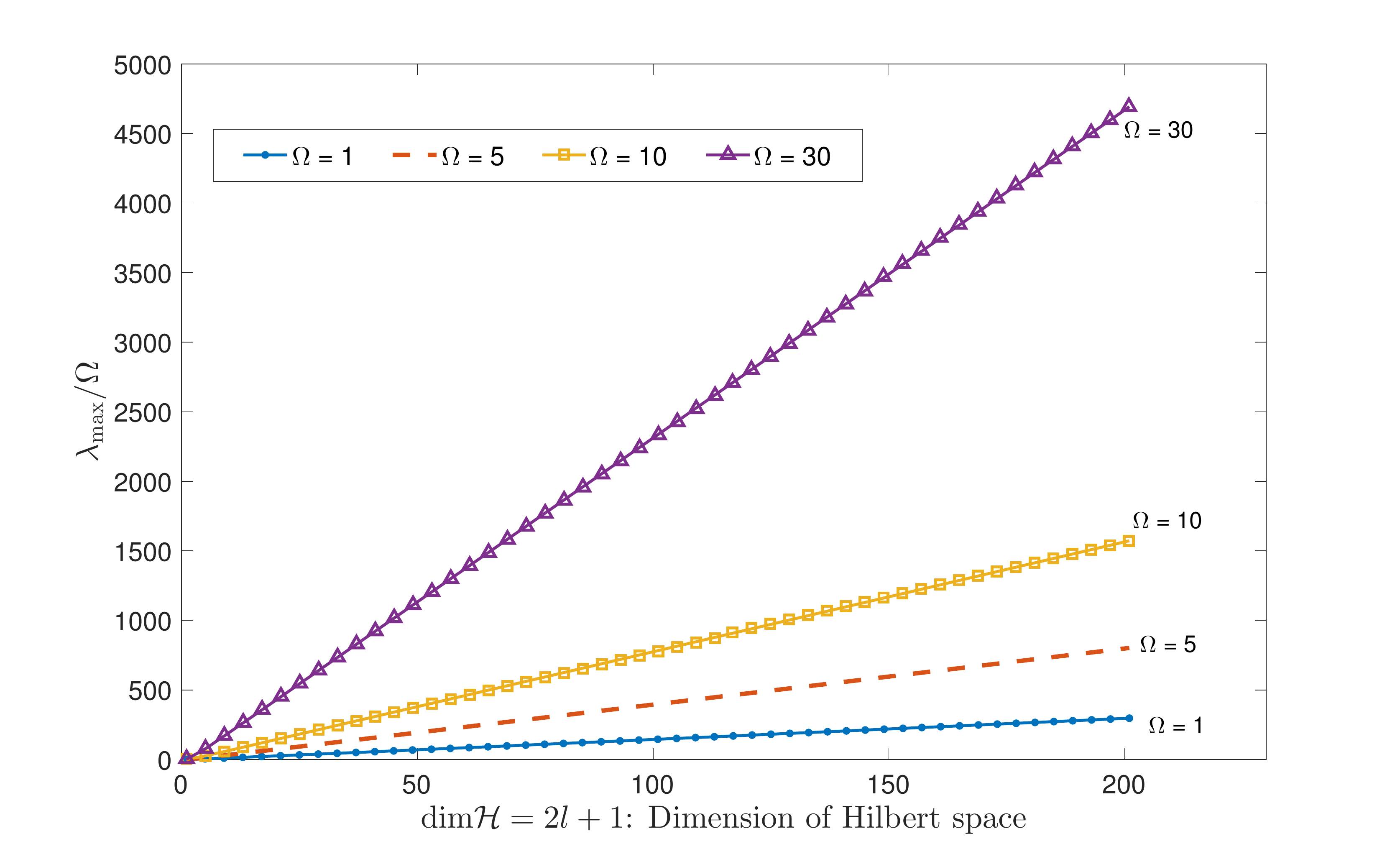}
\caption{Plot of the maximum eigenvalue (normalized by $\Omega$) of the finite-oscillator as a function of dimension $\Dim \hs = 2l + 1$ for different values of $\Omega$. A linear trend is observed.}
\label{oscillator_max}
\end{figure}
Another important feature to consider is the maximum eigenvalue of $\ham$, $\lambda_{\mathrm{max}}$. While there is no maximum eigenvalue in the infinite-dimensional case, we find that $\lambda_{\mathrm{max}}$ has almost linear behavior in the dimension $\Dim \hs$ of Hilbert space, as plotted in Figure (\ref{oscillator_max}).

 A bound for $\lambda_{\mathrm{max}}$ can easily be given,
\begin{equation}
\lambda_{\mathrm{max}} \leq \frac{1}{2} \left( 1 + \Omega^{2} \right) \left(l \alpha \right)^{2} = \frac{\pi l^{2}}{2l + 1} \left(1 + \Omega^{2}   \right)\: ,
\end{equation}
where we have used the fact that for hermitian matrices $P, Q $ and $R$ such that $P = Q + R$, the maximum eigenvalue of $P$ is at most the sum of maximum eigenvalues of $Q$ and $R$.

At the other end, while the minimum eigenvalue, normalized by $\Omega$, has a constant $1/2$ value for the vanilla, infinite-dimensional oscillator, we find a richer structure for the minimum eigenvalue of the finite oscillator, plotted in Figure (\ref{oscillator_min}). This is itself a reflection of the non-centrality of the commutator $\left[\hat{a},\hat{a}^{\dag} \right] \neq 1$, and we see how the lowest eigenvalue normalized by $\Omega$ is suppressed for larger values of $\Omega$ for a given Hilbert space. 
\begin{figure}[h]
\includegraphics[width=\textwidth]{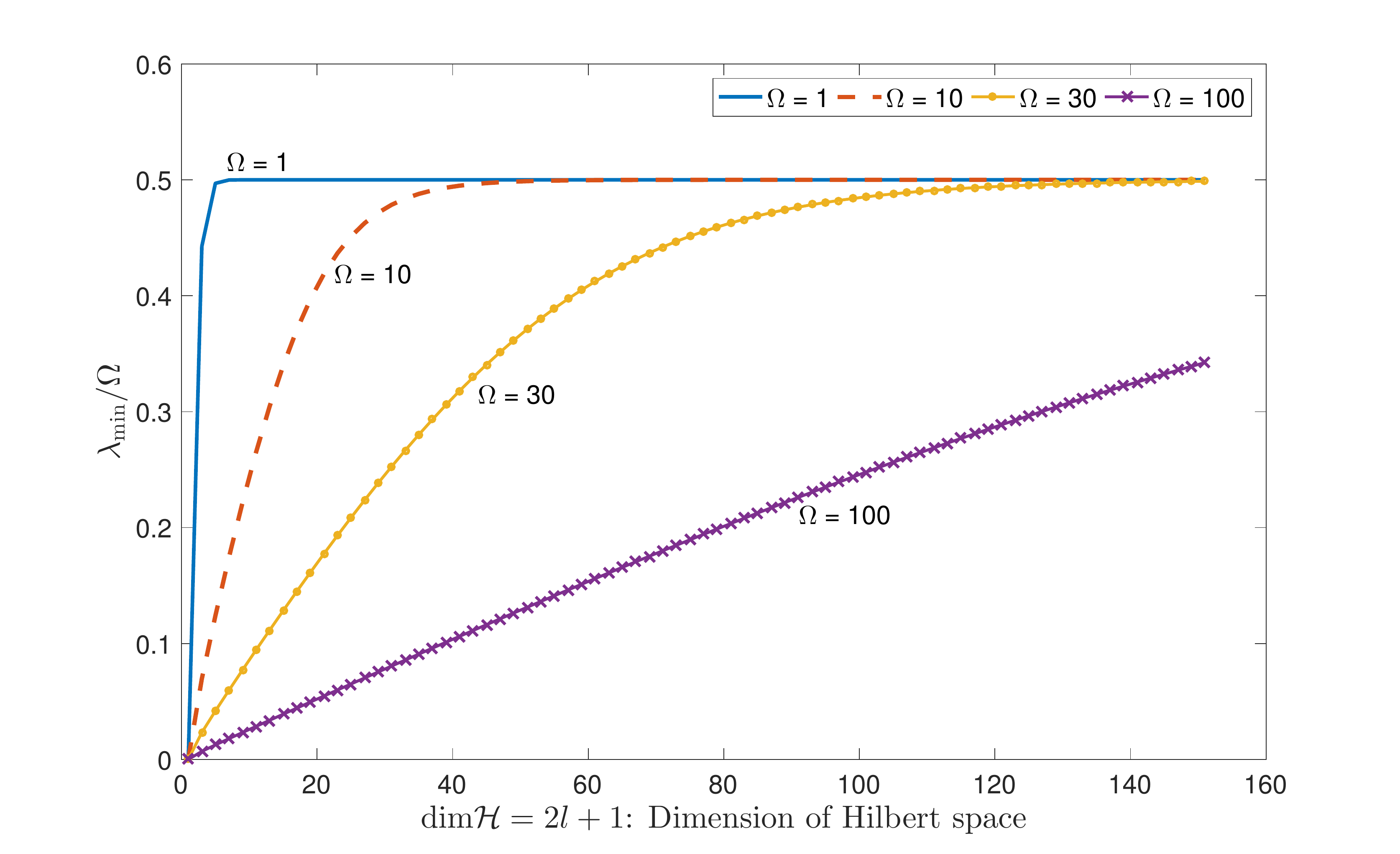}
\caption{Plot of the minimum eigenvalue(normalized by $\Omega$) of the finite-oscillator as a function of dimension $\Dim \hs = 2l + 1$ for different values of $\Omega$. A gradual build-up is noticed, with more suppression for larger $\Omega$ for a given $\hs$, which saturates to the vanilla, infinite-dimensional result of $\lambda_{\mathrm{min}}/ \Omega = 0.5$.}
\label{oscillator_min}
\end{figure}
These features of the finite-dimensional oscillator spectrum could play a crucial role in the physics of locally finite-dimensional models of quantum gravity.

\section{Discussion}
\label{sec:GCA_discussion}

Quantum-mechanical models have been extensively studied in both finite- and infinite-dimensional Hilbert spaces; the connection between the two contexts is less well-understood, and has been our focus in this paper.
Infinite-dimensional models are often constructed by quantizing classical systems that have a description in terms of phase space and conjugate variables.
We have therefore studied the Generalized Pauli operators as a tool for adapting a form of conjugate variables to the finite-dimensional case, including the appropriate generalization of the Heisenberg canonical commutation relations.

An advantage of the GPO is that it is completely general, not relying on any pre-existing structure or preferred algebra of observables.
This makes it a useful tool for investigating situations where we might not know ahead of time what such observables should be, such as in quantum gravity.
We have investigated the development of position/momentum variables, and an associated notion of operator collimation, within this framework.
This analysis revealed hints concerning the special nature of the true Hamiltonian of the world, especially the distinction between position and momentum and the emergence of local interactions (and therefore of space itself).

As we have seen, features of a theory based on an intrinsic finite-dimensional Hilbert space can be very different than one based on naive truncation of an infinite-dimensional one. This is particularly seen in the example of the finite-dimensional quantum harmonic oscillator discussed in Section \ref{sec:finite_models}, where the spectrum of the oscillator differs from a simple truncation of the vanilla spectrum based on the infinite-dimensional oscillator. A consistent finite-dimensional construction applied to field theory could have important consequences for issues such as the hierarchy problem, the cosmological constant problem, and Lorentz violation, and may lead to corrections in Feynman diagrams for given scattering problems. In addition to its possible role in field theory, modifications to the commutation relation of conjugate variables (departure from it being a commuting number) can further lead to modifications to uncertainty relations. It has been shown \citep{Maggiore:1993kv, Maggiore:1993rv, Bosso:2017hoq, PhysRevD.94.123505} (and references therein) that taking into account gravitational effects will lead to modified commutation relations, and the GPOs can provide a natural way to understand these in terms of the local dimension of Hilbert space in a theory with gravity. The GPO can also play an important role in our understanding of emergent classicality in a finite-dimensional setting, where in some preferred factorization of Hilbert space into sub-systems, the conjugate variables can be identified as classical conjugates such as positions and momenta.

Constructions based on the GPO have also been shown to be important in quantum error correction and fault tolerance \citep{Gottesman:2000di}, where one can further try and quantify robustness of different operators based on a notion of operator collimation. Once dynamics is added to the problem, one can study the operator collimation of operators as a function of time, understanding how their support on Hilbert space evolves, and this can be connected with ideas in quantum chaos and out-of-time-ordered-correlators (OTOCs) \citep{Maldacena:2015waa}.

In future work we plan to further explore the emergence of spacetime and quantum field theory in a locally finite-dimensional context.

\section*{Acknowledgements}
We would like to thank Anthony Bartolotta, ChunJun (Charles) Cao, Aidan Chatwin-Davies, Swati Chaudhary, Prof. R. Jagannathan and Jason Pollack for helpful discussions during the course of this project. This research is funded in part by the Walter Burke Institute for Theoretical Physics at Caltech and by DOE grant DE SC0011632.

\bibliographystyle{utphys}
\bibliography{GCA_finite_dim}

\end{document}